\documentclass[12pt]{article}

\usepackage{amsmath, setspace}
\usepackage{amssymb, xspace}

\usepackage{graphicx}
\usepackage{color} 

\usepackage{subcaption}
\usepackage{hyperref}
\usepackage{times}
\usepackage{cite}
\usepackage[table,rgb]{xcolor}
\usepackage{multirow}
\usepackage[misc,geometry]{ifsym}

\usepackage{caption}
\DeclareCaptionLabelSeparator{pipe}{$\; | \;$}
\definecolor{abstarctBlue}{rgb}{0.0706, 0.349, 0.6667} 
\definecolor{abstarctBlue2}{cmyk}{ 0.5143,   0.2857,    0.0286,    0.3714}
\definecolor{Section}{cmyk}{0.2,0.8,0.8,0.3}

\usepackage{soul}
\sodef\an{\fontfamily{phv}\selectfont}{.08em}{1em plus1em}{0.5em plus.1em minus.1em} 
\sodef\ann{\fontfamily{phv}\selectfont}{0.04em}{0.5em plus0.02em}{0.1em plus.1em minus.1em}

\makeatletter
\renewcommand{\@biblabel}[1]{\quad#1.}
\makeatother

\date{}

\usepackage{overpic}
\newcommand*{\hvfont}{\fontfamily{phv}\selectfont}

\newcommand{\inb}[2]{ \begin{overpic}[width = .42\textwidth]{#1} \put(0,75){\large \bf \hvfont #2}\end{overpic} }

\newcommand{\inc}[2]{ \begin{overpic}[width = .4\textwidth]{#1} \put(0,79){\large \bf \hvfont #2}\end{overpic} }

\newcommand{\fg}{\textcolor{linkcolor}{Fig.}~\ref}
\newcommand{\supp}{Methods\xspace	} 

\newcommand{\blue}{\textcolor{blue}	}

\newcommand{\pana}{({\bf a})\xspace}
\newcommand{\panb}{({\bf b})\xspace}
\newcommand{\panc}{({\bf c})\xspace}
\newcommand{\pand}{({\bf d})\xspace}
\newcommand{\pane}{({\bf e})\xspace}

\newcommand{\figs}[1]{{Fig.~S#1}} 

\definecolor{citecolor}{rgb}{0.071, 0.36, 0.67}   
\definecolor{linkcolor}{rgb}{0.071, 0.4, 0.67}  
\hypersetup{
colorlinks=true, 
citecolor=citecolor,
linkcolor=linkcolor,
urlcolor=linkcolor
}

\newcommand{\name}{SCoPE-MS\xspace	} 
\newcommand{\rna}{mRNA\xspace	}

\graphicspath{ 
{Figs/} 	   
}

\let\citep=\cite
\let\citet=\cite

\def \concept {{\bf Validating \name by classifying single cancer cells based on their proteomes.}
\pana Conceptual diagram and work flow of \name. Individually picked live cells are lysed by sonication, the proteins in the lysates are digested with trypsin, the resulting peptides labeled with TMT labels, combined and analyzed by LC-MS/MS (Orbitrap Elite).     
\panb Design of control experiments used to test the ability of \name to distinguish U-937 cells from Jurkat cells. Each set was prepared and quantified on a different day to evaluate day-to-day batch artifacts. 
\panc Unsupervised principal component (PC) analysis using data for quantified proteins from the experiments described in panel (b) stratifies the proteomes of single cancer cells by cell type. Protein levels from 6 bulk samples from Jurkat and U-937 cells are also projected and marked with filled semitransparent circles.  The two largest PCs explain over $50\%$ of the variance. Similar separation of Jurkat and U-937 cells is observed when different carrier cells are used, \figs{2}; see \hyperref[note1]{\bf Supplementary note 1}.  
\pand Distributions of protein levels across single U-937 and Jurkat cells indicate cell-type-specific protein abundances. 
\pane Adenocarcinoma cells (MDA-MB-231) expressing mCherry and LifeAct-iRFP670 were sorted by Aria FACS  into a 96 well plate, one cell per well. The relative levels of mCherry and iRFP were estimated by the sorter (from their florescence intensity) and by SCoPE-MS, and the two estimates compared by their Spearman correlations $(\rho)$. 
}

\def \prot  {{\bf Identifying protein covariation across differentiating ES cells.}\\ 
\pana   Clustergrams of pairwise protein-protein correlations in cells differentiating for  3, 5, and 8 days after LIF withdrawal. The correlation vectors were hierarchically clustered based on the cosine of the angles between them. All single cell sets used the same carrier channel which was comprised of cells mixed from different time points.     
\panb All pairwise Pearson correlations between ribosomal proteins (RPs) were computed by averaging across cells. The correlation matrix was clustered, using the cosine between the correlation vectors as a similarly measure. 
\panc To evaluate the similarity in the relative levels of functionally related proteins, we computed the Pearson correlations within sets of functional related proteins as defined by the gene ontology (GO). These sets included protein complexes, lineage-specific proteins, and proteins functioning in cell growth and division. The distribution of correlations for all quantified proteins is also displayed and used as a null distribution. The null distribution has a positive bias, i.e.,  covariance component common to all proteins, perhaps reflecting  contributions of batch effects.  
 }

\def \pca  {{\bf Principal component analysis of differentiating ES cells.}\\ 
\pana   Distributions of protein abundances for all proteins quantified from $10^7$ differentiating ES cells \citep{Dynamic_PTR} or in at least one single-cell \name set at FDR $\le 1\; \%$.  The probability of quantifying a protein by \name is close to $100\; \%$ for the most abundant proteins quantified in bulk samples and decreases with protein abundance, for total of 1526 quantified proteins.     
\panb The proteomes of all single EB cells were projected onto their PCs, and the marker of each cell color-coded by day. The single-cell proteomes cluster partially based on the days of differentiation.
\panc A tabular display of the variance explained by the principal components from panel c and their correlations to the days of differentiation and the missing data points for each cell. 
({\bf d, e}) The proteomes of cells differentiating for 8 days were projected onto their PCs, and the marker of each cell color-coded based on the normalized levels of all proteins from the indicated gene-ontology groups.
 }

\def \rnaProt  { {\bf Coordinated mRNA and protein covariation in differentiating ES cells.}\\
\pana Clustergram of pairwise correlations between mRNAs with 2.5 or more reads per cell as quantified by inDrop in single EB cells \citep{klein2015droplet}. 
\panb Clustergram of pairwise correlations between proteins quantified by \name in 12 or more single EB cells. 
\panc The overlap between corresponding RNA from (a) and protein clusters from (b) indicates similar clustering patterns. 
\pand Protein-protein correlations correlate to their corresponding mRNA-mRNA correlations. Only genes with significant mRNA-mRNA correlations were used for this analysis.    
\pane The concordance between corresponding mRNA and protein correlations (computed as the correlation between between corresponding correlations \citet{Slavov_2009}) is high for ribosomal proteins (RPL and RPS) and lower for developmental genes; distribution medians are marked with red pluses.  
Only the subset of genes quantified at both RNA and protein levels were used for all panels. 
 }

\def \noise  {{\bf Contribution of background noise to quantification of peptides in single cells.}
\pana Reporter ion (RI) intensities in a SCoPE set in which the single cells were omitted while all other steps were carried out, i.e., trypsin digestion, TMT labeling and addition of carrier cells in channel 131. Thus, RI intensities in channels $126 - 130C$ correspond to background noise. The distribution of RI intensities in the inset shows that the RI for most peptides in channels $126 - 130C$ are zero, i.e., below the MaxQuant noise threshold. The y-axis is limited to 150 to make the mean RI intensities visible. The mean RI intensity for single-cell channels is about 500.  
\panb Mean RI intensities for a TMT set in which only 6 channels contained labeled proteome digests and the other 4 were left empty. Channels 126, 127N, 128C, and 129N correspond to peptides diluted to levels corresponding to 100, 100, 200 and 300 picograms of cellular proteome, channel 131 corresponds to the carrier cells (bars truncated by axes), and the remaining channels were left empty. The RI for most peptides are not detected in the empty channels, and their mean levels very low. This suggests that background noise is low compared to the signal from peptides corresponding to a single cell.        
\panc The intensities for representative RIs from (b) are plotted, color coded by their mean intensity. The results show that both low (blue) and high (red) abundant RIs exhibit the expected scaling with the increase of the labeled cell lysate.         
 }

\def \carrier_Effect {{\bf Relative quantification is independent from the carrier channel.}    
\pana Design of control experiments used to test the ability of \name to distinguish U-937 cells from Jurkat cells independently from the cells in the carrier channel. The carrier channel in each \name set contained 200 cells: 200 Jurkat cells in set 1, 200 U-937 cells in set 2, and 200 HEK-293 cells in set 3.    
\panb Unsupervised principal component (PC) analysis using data for quantified proteins from the experiments described in panel (a) stratifies the proteomes of single cancer cells by cell type regardless of the type of cells used in the carrier channel. 
\panc To explore the extent to which peptides from the carrier channel might affect relative quantification in the single-cell channels, we computed the pairwise correlations between the relative peptide levels of a \name set. Relative peptide levels were computed by first normalizing the RI intensities in each channel to a median of 1 (to correct for different amount of total protein, especially in the carrier channel) and then by dividing the vector of RI intensities for each peptide by its mean, to remove the large differences in abundances between different peptides. The correlations between these relative estimates indicate that Jurkat cells correlate positively to  Jurkat cells and negatively to U-937 cells. The converse holds for U-937 cells. Importantly, the carrier channel correlates negatively with most single cells, except for a weak correlation with one U-937 cell, perhaps reflecting slightly high contribution of U-937 cells to the carrier channel.       
}

\def \JUratios  {\small {\bf Accuracy of \name quantification.}
\pana Comparison between protein levels estimates from bulk samples and from single cells. The single-cell protein estimates are the average from 12 Jurkat cells from the experiments described in Fig.~1b and equal as the summed up precursor ion areas apportioned by RI intensities.       
\panb Comparison between relative protein levels (fold changes) estimated from bulk samples and from single cells. The single-cell protein estimates are the ratio between average levels summed across 12 Jurkat cells over the average level summed across 12 U-937 cells. Proteins whose bulk estimates did not change between Jurkat and U-937 cells (fold change less than 10\%) were omitted from the plot.        
\panc A correlation matrix of all pairwise Pearson correlations among the ratios of peptide abundances in U-937 and in Jurkat cells from Set 2 in Fig.~1b. The superscripts corresponds to the TMT labels ordered by mass, with 1 being 126, 2 being 127N and so on. The positive correlations among estimates from different combinations of TMT channels suggest good consistency of relative quantification. 
\pand Distributions of correlations between technical replicates of peptide ratios measured in two halves of the same single-cell set; each measurement estimated the peptide ratios from peptides corresponding to 1/2 cell. The first distribution correspond to correlations from across all measured peptides. The other distributions correspond to the correlations computed from the subset of peptides having coefficient of variation (CV) above the indicated percentile, i.e., peptides with larger fold changes. The red crosses mark the distribution medians. Correlations were computed with log transformed protein levels and ratios.         
}

\def \pc1byDay  {{\bf Proteome coverage of differentiating ES cells and distributions of the PC 1 loadings by day of differentiation.}     
\panb The proteomes of all differentiating single cells were decomposed into singular vectors and values, and distributions of the loading (elements) of the singular vector with the largest singular value, i.e., PC 1, shown as violin plots. Individual blue circles correspond to single cells, and the red crosses correspond to the medians for each day.   
 }

\def \RP_corrs  {{\bf Correlations between ribosomal proteins} 
\pana All pairwise Pearson correlations between ribosomal proteins (RPs) were computed by averaging across cells. The correlation matrix was clustered, using the cosine between the correlation vectors as a similarly measure.   
 }

 \def \Sorted_PEPs  {{\bf Spectrum of peptide identification across \name sets} 
The peptides from a few \name sets from Fig.~1 and from Fig.~S3 were rank sorted by the confidence of their identification as quantified by the posterior error probability (PEP).  The rank sorted PEPs for each set are color-coded based on the number of quantified peptides at PEP $< 0.03$, excluding peptides from contaminant proteins. The results exemplify the variability in the number of peptides quantified in different \name sets. For some sets, relaxing the false discovery rate (FDR) to $3\%$ increases significantly the number of quantified peptides while keeping false positives below $3\%$.   
 }

\date{}
\begin{document}

\begin{spacing}{1.6}
\noindent {\Large \bf 
Mass-spectrometry of single mammalian cells quantifies \\ proteome heterogeneity during cell differentiation
}

\end{spacing}
\vspace{10mm}

\noindent\ann{
Bogdan Budnik,$^{1}$ 
Ezra Levy,$^{2}$
Guillaume Harmange,$^{2}$	 and
Nikolai Slavov$^{2,3}$ 
} 

{\small 
\noindent 
$^{1}$MSPRL, FAS Division of Science, Harvard University, Cambridge, MA 02138, USA\\
$^{2}$Department of Biology, Northeastern University, Boston, MA 02115, USA\\
$^{3}$Department of Bioengineering, Northeastern University, Boston, MA 02115, USA\\
}

\thispagestyle{empty}
\vspace{1cm}


\begin{spacing}{1.5} 
\noindent{\bf
Cellular heterogeneity is important to biological processes, including cancer and development. However, proteome heterogeneity is largely unexplored because of the limitations of existing methods for quantifying protein levels in single cells. To alleviate these limitations,  we developed Single Cell ProtEomics by Mass Spectrometry (\name), and \mbox{validated} its ability to identify distinct human cancer cell types based on their proteomes. We used  \name to quantify over a thousand proteins in differentiating mouse embryonic stem (ES) cells. The single-cell proteomes enabled us to deconstruct cell populations and infer protein abundance relationships. Comparison between single-cell proteomes and transcriptomes indicated coordinated mRNA and protein covariation. Yet many genes exhibited functionally concerted and distinct regulatory patterns at the mRNA and the protein levels, \mbox{suggesting} that post-transcriptional regulatory mechanisms contribute to proteome remodeling during lineage specification, especially for developmental genes.  \name is broadly applicable to measuring proteome configurations of single cells and linking them to functional phenotypes, such as cell type and differentiation potentials.                
}
\vspace{1cm}

\newpage 

\section*{Background}
Cellular systems, such as tissues, cancers, and cell cultures, consist of a variety of cells with distinct molecular and functional properties. Characterizing such cellular differences is key to understanding normal physiology, combating cancer recurrence, and enhancing targeted stem cell differentiation for regenerative therapies \citep{dean2005tumour, cohen2008dynamic, semrau2015AnnualReview, Arjun_2016_Noise, Levy_Review_2018}; it demands quantifying the proteomes of single cells.  
 
However, quantifying proteins in single mammalian cells has remained confined to fluorescent imaging and antibodies.  Fluorescent proteins have proved tremendously useful but are limited to quantifying only a few proteins per cell and sometimes introduce artifacts \citep{Levy_Review_2018, landgraf2012segregation}.  Multiple antibody-based methods for quantifying proteins in single cells have been recently developed, including, CyTOF \citep{bandura2009mass, bendall2011single}, single-cell Western blots \citep{hughes2014single}, and Proseek Multiplex, an immunoassay readout by PCR  \citep{darmanis2016simultaneous}. These methods can quantify up to a few dozen endogenous proteins recognized by highly-specific cognate antibodies and have enabled exciting research avenues \citep{Levy_Review_2018}. Still, the throughput and accuracy of antibody-based methods are limited by cellular permeability, molecular crowding,  epitope accessibility, and the availability of highly-specific antibodies that bind their cognate proteins stoichiometrically.

On the other hand, the application of liquid chromatography (LC) and tandem Mass Spectrometry (MS/MS) to bulk samples comprised of many cells allows for the confident identification and quantification of thousands of proteins \citep{Aebersold2003,de2008comprehensive, cox2008maxquant, Slavov_exp,wilhelm14, Dynamic_PTR, nesvizhskii2014proteogenomics}. To develop approaches that may bring at least some of this power of LC-MS/MS to single mammalian cells, we considered all steps of well-established bulk protocols and how they may be adapted to much more limited samples.  We were motivated by the realization that most proteins are present at over 50,000 copies per cell\citep{milo2010bionumbers, schwanhausser2011global_Corrigendum} while modern MS instruments have sensitivity to identify and quantify ions present at 100s of copies \citep{zubarev2013orbitrap, Specht_Perspective_2018}.

Most protocols for bulk LC-MS/MS begin by lysing the cells with detergents or urea \citep{dhabaria2015high}. Since these chemicals are incompatible with MS, they have to be removed by cleanup procedures. These cleanup procedures can result in substantial losses of protein,  and colleagues have developed advanced methods, such as SP3 \citep{hughes2014ultrasensitive} and iST\citep{kulak2014minimal},  that  minimize protein losses and allow for quantifying thousands of proteins from  samples having just a few micrograms of total protein \citep{sielaff2017evaluation, dhabaria2015high}. Indeed, the SP3 method has  been successfully used  for purifying and quantifying proteins from single human oocytes ($\sim 100 \mu m$ diameter) \citep{oocytes2016Krijgsveld}. Still, most mammalian cells  are smaller ($10 - 15 \mu m$ diameter) \citep{milo2010bionumbers}, and we were not confident that we could clean-up their cell lysates (having about $500 pg$ of total protein) without incurring large protein losses. Thus, we sought to obviate cleanup (and therefore eliminate cleanup-related losses) by replacing chemical lysis with mechanical lysis by focused acoustic sonication \citep{dhabaria2015high, Ivanov2015RareCells}.  

Before being ionized and sent for MS analysis,  peptides have to be separated \citep{Aebersold2003,Slavov_exp,wilhelm14}. The separation for bulk samples is usually accomplished by nano liquid chromatography (nLC). To reduce losses due to proteins adhering to the large surface area of nLC columns, low-input samples can also be separated by capillary electrophoresis\citep{lombard2016single}. We sought to minimize nLC losses by mixing  labeled peptides from single cells with labeled carrier peptides so that many of the peptides lost due to nLC adhesion will be carrier peptides rather than single-cell peptides. This strategy deviates from standard protocols for bulk LC-MS/MS.

Once injected into an MS instrument, peptide ions need at least two rounds of MS analysis for confident sequence identification\citep{eng1994approach, cox2008maxquant, sinitcyn2018computational}. The first MS scan (MS1) determines the mass over charge ratio (M/z) for ions that entered the instrument. 
Then, selected ions are accumulated, fragmented, and their fragments analyzed by an MS2 scan\citep{Aebersold2003,sinitcyn2018computational}. The most commonly used fragmentation methods break peptides at the peptide bonds with efficiency that varies much from bond to bond\citep{sinitcyn2018computational}. Since some fragments are produced with low efficiency, they will not be detected if the peptide ions have low abundance; if not enough fragments are detected, the peptide cannot be sequenced. We sought to alleviate this limitation by sending for MS2 analysis labeled peptide ions having the same M/z (and thus the same sequence labeled with sample-specific barcodes) from multiple single cells and from carrier cells so that a larger number of peptide ions are fragmented and used for sequence identification. This strategy is built upon the foundational ideas of isobaric tandem mass tags (TMT) \citep{thompson2003tandem, Pappin2004multiplexed, sinitcyn2018computational}. TMT labels are used with conventional  bulk LC-MS/MS to label samples of equal total protein amount\citep{Slavov_ribo, Slavov_exp, sinitcyn2018computational} and offer many advantages, albeit quantification can be affected by ion co-isolation \citep{savitski2013measuring}; our implementation of TMT, as described below, uses a carrier channel with much higher total protein abundance than the single cells and deviates from the standard protocols.

MS instruments have expanding but limited capacity for parallel ion processing and analysis \citep{Aebersold2003,michalski2011mass, meier2015parallel}. Thus increase in throughput has been driven in part by decreasing the time for each step, reaching low millisecond ranges for MS scans and for ion accumulation for bulk LC-MS/MS analysis \citep{michalski2011mass, Slavov_exp}. On the other hand, nLC elution peaks have widths on the order of seconds \citep{Ivanov2015RareCells, Specht_Perspective_2018}. Thus, if a peptide elutes from the nLC for 8s and is accumulated (sampled) for only 50ms by an MS instrument, the instrument will measure only a small fraction of the peptide molecules in the sample\citep{Specht_Perspective_2018}. This inefficient sampling is compensated  for in  standard bulk methods by the large input amount but becomes problematic for low-input samples; counting noise alone can undermine quantification\citep{Specht_Perspective_2018}. In this work, we sought to alleviate the sampling limitation by increasing the ion accumulation (sampling) time at the expense of quantifying fewer peptides per unit time. We have discussed additional strategies for increasing sampling and mitigating its trade-offs in a recent perspective\citep{Specht_Perspective_2018}.

\section*{Results}
Thus, to develop a high-throughput method for Single Cell ProtEomics by Mass Spectrometry (\name), we had to alter substantially the LC-MS/MS  methods for bulk samples. In particular, we had to resolve two major challenges: (i) delivering the proteome of a mammalian cell to a MS instrument with minimal protein losses and (ii) simultaneously identifying and quantifying peptides from single-cell samples.   
To overcome the first challenge, we manually picked live single cells under a microscope and lysed them mechanically (by Covaris sonication in glass microtubes), \fg{concept}a. This method was chosen to obviate chemicals that may undermine peptide separation and ionization or sample cleanup that may incur significant losses. The proteins from each cell lysate were quickly denatured at $90\; ^oC$ and digested with trypsin at $45\; ^oC$ overnight, \fg{concept}a. Special care was taken to ensure that each tube contained only one cell. See \supp for full experimental details. 

\newcommand{\fpa}[2]{ \begin{overpic}[width = .3\textwidth]{#1} \put(-5,70){\large \bf \hvfont  #2}\end{overpic} }
\newcommand{\inh}[2]{ \begin{overpic}[width = .28\textwidth]{#1} \put(-5,88){\large \bf \hvfont  #2}\end{overpic} }
\begin{figure}[h!]
   \begin{overpic}
   		[width = .98\textwidth]{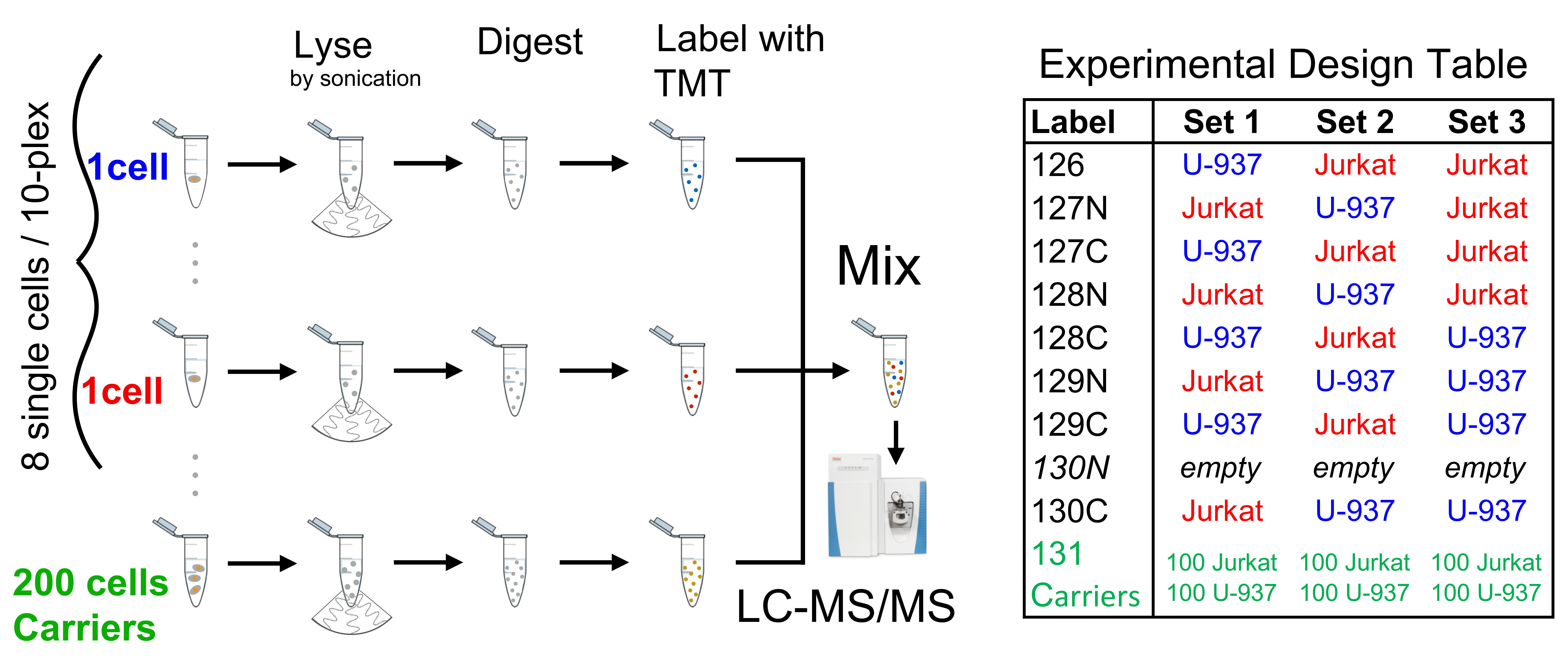} 
   		\put(-1,  38 ){\large \bf \hvfont  a}
   		\put(61,  38 ){\large \bf \hvfont  b}
   	\end{overpic}
   	 \\ [2em]
	\inh{PCA_U_J_12.pdf}{c} 
	\hspace{0.02\textwidth}
	\inh{Dist_U_J_4}{d} 
	\hspace{0.02\textwidth}
	\inh{FACS.pdf}{e} 
 	\caption{\concept}
 	\label{concept}
\end{figure}

To overcome the second challenge, we made novel use of tandem mass tags (TMT). This technology was developed for multiplexing\citet{thompson2003tandem, Pappin2004multiplexed}, which is usually employed for cost-effective increase in throughput. Even more crucial to our application, TMT allows quantifying the level of each TMT-labeled peptide in each sample while identifying its sequence from the total peptide amount pooled across all samples\citet{thompson2003tandem, Pappin2004multiplexed}. \name capitalizes on this capability by augmenting each single-cell set with a sample comprised of $\sim 100-200$ carrier cells that provide enough ions for peptide sequence identification, \fg{concept}a.  The carrier cells also help with the first challenge by reducing losses from single cells, since most of the peptides sticking to tips and tube walls originate from the carrier cells. Thus, the introduction of labeled carrier cells into single-cell TMT sets helps overcome the two major challenges.

Quantification of TMT-labeled peptides relies on reporter ions (RI) whose levels reflect both peptide abundances and noise contributions, such as coisolation interference and background noise\citet{Pappin2004multiplexed, savitski2013measuring}. 
The low protein abundance poses extreme challenges to the signal-to-noise ratio (SNR) and requires careful evaluation even of aspects that are well established and validated in bulk MS measurements. 
 To evaluate the contribution of background noise to single-cell RI quantification, we estimated the signal-to-noise ratio (SNR), \figs{1}. The estimates indicated that RI intensities are proportional to the amount of labeled single-cell proteomes, and very low for channels left empty. These data suggest that the signal measured in single cells exceeds the background noise by 10-fold or more. As an added SNR control for every TMT set, \name leaves the 130N channel empty, so that 130N RI reflect both isotopic cross-contamination from channel 131 and the background noise. We further verified that RI intensities in a channel are proportional to the protein amount labeled in that channel for both lowly and highly abundant RIs, \figs{1}b,c.

To evaluate the ability of \name to distinguish different cell types, we prepared three label-swapped and interlaced TMT sets with alternating single Jurkat and U-937 cells, two blood cancer cell lines with average cell diameter of only $11\; \mu m$ (\fg{concept}b). The levels of all 767 proteins quantified in single cells were projected onto their principal components (PC). The two-dimensional projections of single-cell proteomes clustered by cell type and in proximity to the projection of bulk samples from the same cell type (\fg{concept}c), suggesting that \name can identify cell types based on their proteomes. This cell-type stratification is not driven just by highly abundant proteins since the mean levels of each protein across the single cells was set to one; thus highly and lowly abundant proteins contributed equally to cell clustering. We observed similar cell-type clustering when using carrier cells from different types, affirming the expectation that the relative quantification of proteins in single cells is not contaminated by the carrier cells, \figs{2}; see \hyperref[note1]{\bf Supplementary notes 1 and 2}.    
To further test the quantification of cell-type specific protein expression, we identified proteins whose levels vary less within a cell type than between cell types. We found 107 proteins showing such trends at FDR $< 2\%$; see representative distributions for such proteins in \fg{concept}d. 

Next, we sought to compare \name quantification against an orthogonal and reliable method for quantifying proteins in single cells, the fluorescence of mCherry and iRFP. To this end, the relative levels of the two proteins were quantified in each single cell by a Fluorescence-Activated Cell Sorting (FACS) sorter  and by \name, \fg{concept}e. The Spearman correlations for both proteins exceed $0.7$, suggesting that estimates of relative protein levels by \name are comparable to those derived by FACS.

Given the difficulty of measuring extremely low protein levels, we further evaluated \name data by comparing the mean estimates across single cells from \fg{concept}b to the corresponding estimates from bulk samples for both absolute (\figs{3a}) and relative (fold-change; \figs{3b}) protein levels. The correlations between bulk and single-cell estimates indicate good agreement despite the noise inherent in single cell measurements. The relative quantification by \name was further evaluated by correlating protein fold-changes estimated from different pairs of Jurkat and U-937 cells labeled with different TMT tags, demonstrating  good consistency of relative quantification for all cells and TMT tags (mean correlation $\rho > 0.5$; \figs{3c}). To eliminate the contribution of biological variability and estimate the reproducibility of the MS measurement alone, we split a \name set in two and quantified each half separately. Comparisons of corresponding protein ratios estimated from each half indicated reliability between $60\%$ and $85 \%$ depending on the magnitude of the fold changes, \figs{3d}. This reliability is achieved with about 70 ng of total protein per cell proteome  \citep{milo2010bionumbers} and compares favorably to reliability for bulk datasets \citep{Franks2016PTR}.  
Taken together, these estimates of quantification accuracy and reproducibility demonstrate that while \name measurements are noisier than bulk MS measurements, they are accurate and reproducible, especially for larger fold-changes.

\subsection*{Protein covariation across differentiating ES cells}
Using \name, we quantified single-cell proteome heterogeneity and dynamics during ES cell differentiation.  To initiate differentiation, we withdrew leukemia inhibitor factor  (LIF) from ES cell cultures and transitioned to suspension culture; LIF withdrawal results in complex and highly heterogeneous differentiation of epiblast lineages in embryoid bodies (EB). We used \name to quantify over a thousand proteins at FDR $= 1\;\%$, and their pair-wise correlations (averaging across single cells) in days 3, 5, and 8 after LIF withdrawal (\fg{prot}a). Cells from different days were processed together to minimize batch biases\citep{irizarry2015widespreadBiases}.  

\begin{figure}[h!]
   		\inh{EBs_Corrs_day3_knn2}{a}   \hspace{0.02\textwidth}
	 	\inh{EBs_Corrs_day5_knn2}{}  \hspace{0.02\textwidth}
	  	\inh{EBs_Corrs_day8_knn2}{} 
	\\  [2em]
	   \inh{RP_corrs_3.pdf}{b} \hspace{0.02\textwidth}
   		\begin{overpic}
   			[width = .65\textwidth]{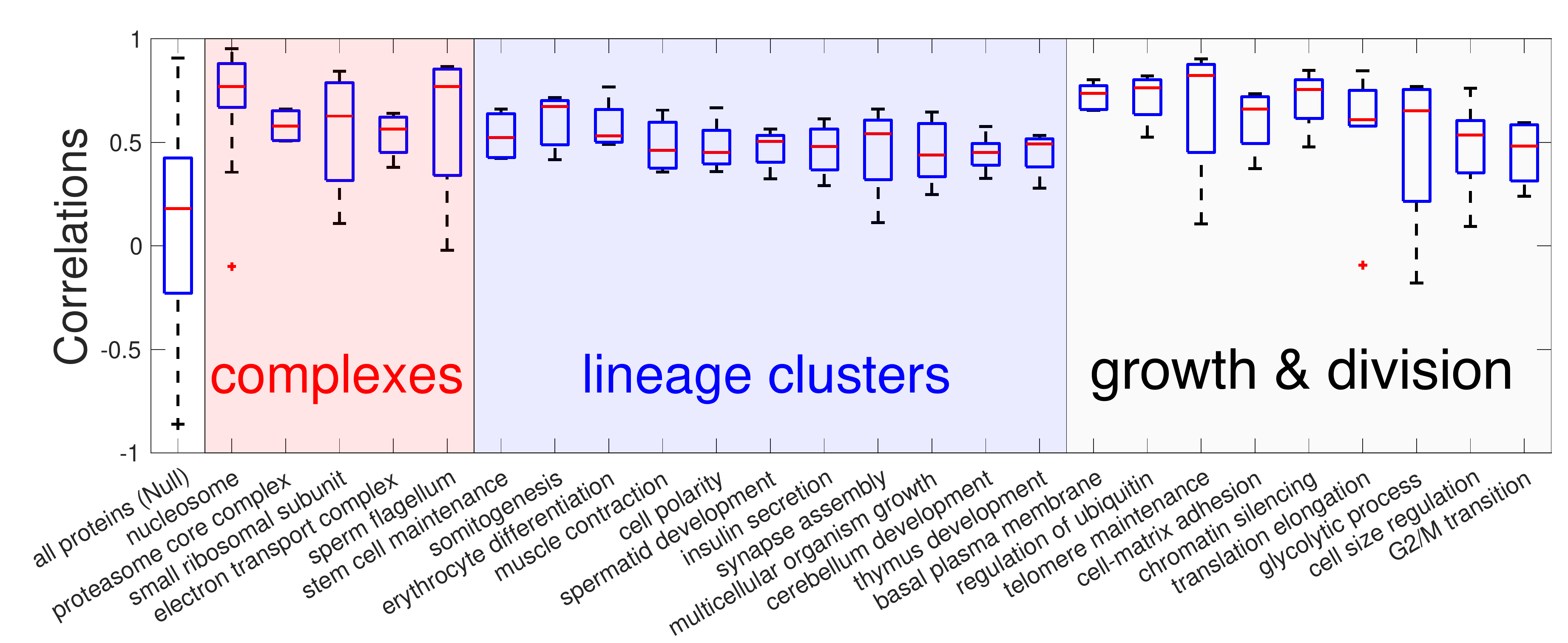} 
   			\put(-1,  38 ){\large \bf \hvfont  c}
   		\end{overpic}
	\caption{\prot}
		\label{prot}
\end{figure} 
 
To  explore the protein covariation across the differentiating single cells, we computed and clustered all pairwise protein-protein correlations, \fg{prot}a.  As cells differentiated and became more distinct from each other, so did the clusters of correlation vectors. Gene set enrichment analysis of the clusters  indicated that  functionally related proteins are over-represented. As expected, proteins forming protein complexes are strongly correlated to each other. For example, most ribosomal proteins (RPs)  correlate positively to each other, \fg{prot}b. A small subset of RPs covary as  a distinct cluster in the bottom right corner of  \fg{prot}b, and this might reflect ribosome specialization, i.e., variation among the RP stoichiometry across the cell lineages that contributes to specialized translation functions \citep{mauro2002ribosome_filter, preiss_2015, Emmott_Ribo_2018}. Alternatively, the cluster might reflect extra-ribosomal functions \citep{wool1996extraribosomal}, and these possibilities need to be evaluated more directly with isolated ribosomes \citep{Slavov_ribo, Emmott_Ribo_2018}. The subunits from other complexes, e.g., the proteasome and the electron transport complex,  also covary as indicated by the positive correlations within these complexes, \fg{prot}c. Similar pattern of covariation is observed for sets of lineage-specific proteins, including proteins with functions specific to neuronal, blood, and muscle cell, \fg{prot}c. Proteins functioning in mRNA translation, metabolism and cell division also covary, most  likely reflecting differences in cell growth and division among the single cells as they differentiate and slow their growth rate.


\subsection*{Principal component analysis of differentiating ES cells}
Next, we sought to classify single cells using all proteins identified and quantified by \name in single ES and EB cells. The quantified proteins tend to be abundant, mostly above the median of $50-100$ thousand copies per / cell, \fg{pca}a. This is expected given that we used shot-gun MS, but combining improvements in \name and targeted MS approaches will enable quantifying substantially less abundant proteins \citep{Specht_Perspective_2018}.  
We projected the proteomes of single cells from all days (190 cells) onto their PCs, \fg{pca}b. The cells partially cluster by time of differentiation; indeed, the loadings of the first three PCs  correlate to the days post LIF withdrawal, \fg{pca}c. However, the clustering by time of differentiation is incomplete, at least in part because of asynchrony in the differentiation \citet{klein2015droplet}.

 \begin{figure}[h!]
	   \begin{overpic}[width = .42\textwidth]{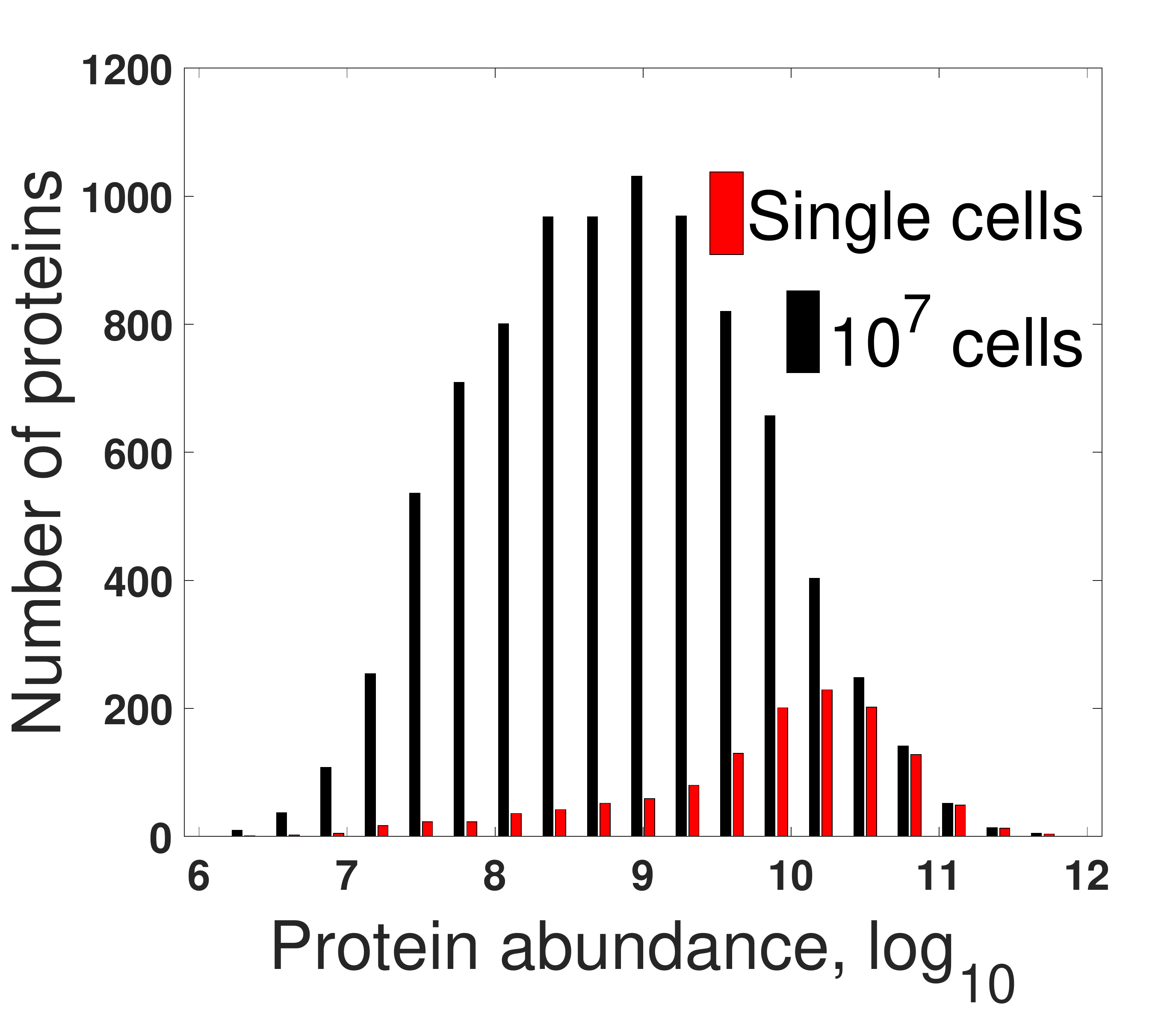} 
	   			\put(-5,80){\large \bf \hvfont  a}
	   			\put(48,74){\large \hvfont {\blue{\small $\Downarrow 10^4$ copies / cell}}} 
	   \end{overpic} 
	   \hspace{0.05\textwidth}
	   \inb{PCA_EBs_Days}{b}   \hspace{0.02\textwidth} 
		\\  [2em]
	   \inh{PCA_PC_table_2.pdf}{c}
		\inh{PCA_EBs_Day8_2.pdf}{d}  \hspace{0.02\textwidth}
		\inh{PCA_EBs_Day8_4.pdf}{e}
	\caption{\pca}
		\label{pca}
\end{figure} 
 
Similar to single-cell RNA-seq, \name did not quantify each gene in each cell. The number of genes with missing quantification varies from cell to cell for single-cell RNAseq methods and this variation is one of the primary sources of variance in the estimated RNA levels \citep{irizarry2015widespreadBiases}. To test if this is the case for \name, we computed the fraction of proteins with missing data for each cell and correlated that fraction to the PCs. The correlations shown in  \fg{pca}c suggest that the degree of  missing data contributes to the variance but less than what has been described for some RNA datasets  \citep{irizarry2015widespreadBiases}.  The degree of missing data can be substantially reduced by using targeted MS  \citep{Specht_Perspective_2018} or its influence mitigated by simply filtering out the proteins with the most missing data or perhaps by more sophisticated normalization approaches. Since the mechanisms generating missing data differ between RNAseq and \name, we expect that the effects of missing data and their management will be different as well.

The  clusters of lineage-specific proteins in \fg{prot} suggest that we have quantified proteomes of distinct cell lineages; thus we attempted to identify cell clusters by projecting the proteomes of cells from day 8 onto their PCs and identifying sets of proteins that are concertedly regulated in each cluster, \fg{pca}d, e. The projection resulted in clusters of cells, whose identity is suggested by the dominant proteins in the singular vectors. We identified biological functions over-represented\citep{Franks2016PTR} within the distribution of PC loadings and colorcoded each cell based on the average levels of proteins annotated to these functions.     
These results suggest that \name data can meaningfully classify cell identity for cells from complex and highly heterogeneous populations.

\subsection*{Coordinated mRNA and protein covariation in single cells }
Klein {\it et al.}\citet{klein2015droplet} recently quantified mRNA heterogeneity during ES differentiation, and we used their inDrop data to simultaneously analyze \rna and protein covariation and to directly test whether genes coexpressed at the \rna level are also coexpressed at the protein level.  
To this end, we computed all pairwise correlations between RNAs (\fg{rnaProt}a) and proteins (\fg{rnaProt}b) for all genes quantified at both levels 
in cells undergoing differentiation for 7 and 8 days. Clustering hierarchically the correlation matrices results in 3 clusters of genes. To compare these clusters, we computed the pairwise Jaccard coefficients, defined as the number of genes present in both classes divided by the number of genes present in either class, i.e., intersection/union). The results (\fg{rnaProt}c) indicate that the largest (green) cluster is 55 \% identical and the medium (blue) cluster is 33 \% identical. This cluster stability is also reflected in a positive correlation between corresponding \rna and protein correlations, \fg{rnaProt}d. The magnitude of this correlation is comparable to protein-\rna correlations from bulk datasets \citep{wilhelm14,Franks2016PTR} and testifies to the quantitative accuracy of both inDrop and \name.

\begin{figure}[h!]
    \fpa{RNA_corrs_day7_2}{a}   \hspace{0.01\textwidth}
	 \fpa{Protein_corrs_day8_2}{b}  \hspace{0.01\textwidth}
	 \fpa{Cluster_Concordance_Jaccard_2}{c} 
	\\ [2em]
   	 \inc{EBs_Corrs_Scatter_d8_knn3.png}{d}
	 \hspace{0.08\textwidth}
	 \inc{EBs_Corr_v_corr_violinP_knn2.png}{e}	
	\caption{\rnaProt}
		\label{rnaProt}
\end{figure} 

Having established a good overall concordance between \rna and protein covariation, we next explored whether and how much this concordance varies between genes with different biological functions. The covariation concordance of a gene was estimated as the similarity of its \rna and protein correlations, using as a similarity metric the correlation between the corresponding correlation vectors as we have done previously \citep{Slavov_2009, silverman2010metabolic}. The median concordance of  ribosomal proteins of both the 60S (RPL) and 40S (RPS) is significantly higher than for all genes, \fg{rnaProt}e. This result indicates that RPL and RPS genes have significantly ($p<10^{-20}$)  more similar gene-gene correlations at the \rna and the protein levels than the other quantified genes. In contrast to RPs, genes functioning in tissue morphogenesis, proteolysis, and development have significantly ($p<10^{-3}$) lower concordance at the \rna and protein level than all genes, \fg{rnaProt}e.  This difference may reflect both differences in the degree of post-transcriptional regulation or measurement noise for the different sets of genes \citep{Franks2016PTR}.

\section*{Discussion}
Until now, the power of LC-MS/MS proteomics has been circumscribed to samples comprised of many cells. Indeed, the TMT manufacturer recommends $100\; \mu g$ of protein per channel, almost $10^6$ more than the protein content of a typical mammalian cell \citep{milo2010bionumbers}. 
 \name bridged this gap by clean sample preparation and by introducing TMT-labeled carrier cells. These innovations open the gates to many further improvements (e.g., increased multiplexing) that will make single-cell MS proteomics increasingly powerful \citep{Specht_Perspective_2018}.

Answering most exciting biological questions demands  quantifying proteins in many thousands of single cells, and we believe that the ideas described and demonstrated here will make such throughput practical and affordable \cite{Specht_Perspective_2018}. At the moment, the cost per cell is about $\$15-30$, but it can be reduced to $\$1-2$ per cell if Covaris tubes are washed and reused and the MS analysis is done on an inhouse MS instrument. We expect that automation and improvements in sample preparation  as well as  increased number of tandem mass tags can reduce the cost well below $\$1$ per cell. Also, the fraction of missing data can be substantially reduced by using targeted MS approaches\cite{Specht_Perspective_2018}.

The floor of protein detecability and quantification with \name (as well as any other bottom-up MS method) depends not only on the abundance of a protein but also on its sequence, i.e., number of peptides produced upon digestion and their propensities to be well separated by the chromatography and efficiently ionized by the electrospray. The implementation of \name in this work, allowed us to quantify mostly abundant proteins present at  $\ge 10^5$ copies / cell and only a few  proteins present at  $\ge 10^4$ copies / cell (those producing the most flyable peptides); see the distribution of abundances of the quantified proteins, shown in \fg{pca}a. However, we are confident that the core ideas underpinning \name can extend the sensitivity to most proteins in a mmamelian cell, down to proteins present at  $\sim 1000$ copies / cell. Such extension requires more efficient delivery of proteins to the MS instruments, and we described specific approaches that can increase the efficiency by orders of magnitude \cite{Specht_Perspective_2018}. These approaches include reduced lysis volume and thus protein loss, and increased sampling of the elution peaks. Such increased sampling is very practical in the context of \name samples analyzed by  MS targetting  proteins of interest, e.g., transcription factors. Since proteins are substantially more abundance than mRNAs, estimates of their abundance is less likely to be undermines by sampling (counting) noise. Thus we believe that building upon this work, future developments in single cell MS have the potential to accurately quantify most proteins in single mammelian cells, including lowly abundant ones\cite{Specht_Perspective_2018}. 

  

\section*{Conclusion}
\name enabled us to classify cells and explore the relationship between \rna and protein levels in single mammalian cells. This first foray into single mammalian proteomes demonstrates that \rna covariation is predictive of protein covariation even in single cells. It further establishes the promise of \name to quantitatively characterize single-cell gene regulation and classify cell types based on their proteomes.  


\bigskip
\bigskip
\bigskip

\noindent {\bf Acknowledgments:} We thank R.~G.~Huffman, S.~Semrau, R.~Zubarev, J.~Neveu, H.~Specht, A.~Phillips and A.~Chen for assistance, discussions and constructive comments, as well as the Harvard University FAS Science Operations for supporting this research project. We thanks A. Raj and B. Emert for the gift of MDA-MB-231 cell line expressing mCherry and LifeAct-iRFP670. 
 This work was funded by startup funds from Northeastern University and a New Innovator Award from the NIGMS from the National Institutes of Health to N.S. under Award Number DP2GM123497. Funding bodies had no role in data collection, analysis, and interpretation. \\

\noindent {\bf Competing Interests:} The authors declare that they have no
competing financial interests.\\
 
\noindent {\bf Contributions:} B.B., and N.S. conceived the research. B.B., E.L., G.H.~and N.S.~performed experiments and collected data; N.S. supervised research, analyzed the data and wrote the manuscript.\\ 

\noindent   {\bf Data Availability:} 		
The raw MS data and the search results were deposited in MassIVE (ID: MSV000082077) and in ProteomeXchange (ID: PXD008985). 
 Supplemental website can be found at: \href{https://web.northeastern.edu/slavovlab/2016_SCoPE-MS/index.html}{northeastern.edu/slavovlab/2016\_SCoPE-MS/}

\newpage
\noindent{\bf \Large
Methods}
\bigskip

\noindent{\bf 
Cell culture}
Mouse embryonic stem cells (E14 $10^{th}$ passage) were grown as adherent cultures in 10 cm plates with 10 ml  Knockout DMEM media supplemented with 10 \% ES certified FBS, nonessential amino acids (NEAA supplement), 2 mM L-glutamine, 110 $\mu M \beta$-mercapto-ethanol, 1 \% penicillin and streptomycin, and leukemia inhibitory factor (mLIF; 1,000 U LIF/ml). 
ES cells were passaged every two days using StemPro Accutase on gelatin coated tissue culture plates. ES differentiation was triggered by passaging the ES cells into media lacking mLIF in low adherence plates and growing the cells as suspension cultures.  
Jurkat and U937 cells were grown as suspension cultures in RPMI medium (HyClone 16777-145) supplemented with 10\% FBS and 1\% pen/strep. Cells were passaged when a density of $10^6$ cells/ml was reached, approximately every two days.

\noindent{\bf 
Harvesting cells for SCoPE-MS}
To harvest cells, embryoid bodies were dissociated by treatment with StemPro Accutase (ThermoFisher \#A1110501 ) and gentle pipetting. Cell suspensions of differentiating ES cells, Jurakt cells or U-937 cells were pelleted and washed quickly with cold phosphate buffered saline (PBS). The washed pellets were diluted in PBS at 4 $^o C$. The cell density of each sample was estimated by counting at least 150 cells on a hemocytometer, and an aliquot corresponding to 200 cells was placed in a Covaris microTUBE-15, to be used for the carrier channel. For picking single cells, two $200\;  \mu l$ pools of PBS were placed on a cooled glass slide. Into one of the pools, $2\;  \mu l$ of the cell dilution was placed and mixed, to further dilute the solution. A single cell was then picked under a microscope into a micropipette from this solution. Then, to verify that only one cell was picked, the contents of the micropipette were ejected into the other pool of PBS, inspected, then taken back into the pipette and placed in a chilled Covaris microTUBE-15. Cell-samples in Covaris microtubes were frozen as needed before cell lysis and labeling.

\noindent{\bf 
Sorting cells by FCAS}
Adenocarcinoma cells (MDA-MB-231) expressing mCherry and LifeAct-iRFP670 were sorted by Aria FACS  into PCR strip-tubes, one cell per tube. Each tube contained $2\: \mu l$ of water and had a max volume of $200\:  \mu l$. The fluorescence of each protein was measured  and the protein abundance estimated after compensation  for the spectral overlap between mCherry and iRFP.

\noindent{\bf 
Cell lysis and digestion}
Each sample -- containing a single cell or carrier cells -- was lysed by sonication in a Covaris S220 instrument (Woburn, MA) \citep{Ivanov2015RareCells}. Samples were sonicated for 180s at 125 W power with 10\% peak duty cycle, in a degassed water bath at 6 $^o C$. During the sonication, samples were shaken to coalesce droplets and bring them down to the bottom of the tube. After lysis, the samples were heated for 15 min at 90 $^o C$ to denature proteins. Then, the samples were spun at 3000 rpm for 1 min, and ($50\:  ng/ \mu l $) trypsin was added;  $0.5\:  \mu l$ to single cells and  $1 \: \mu l$ to carrier cells. The samples were digested overnight, shaking at 45 $^o C$. Once the digest was completed, each samples was labeled with  $1\: \mu l$ of 85mM TMT label (TMT10 kit, Thermo-Fisher, Germany). The samples were shaken for 1 hour in a tray at room temperature. The unreacted TMT label in each sample was quenched with $0.5\:  \mu l$ of 5\% hydroxylamine for 15 min according to manufacturer’s protocol. The samples corresponding to one TMT10 plex were then mixed in a single glass HPLC vial and dried down to $10\:  \mu l$ in a speed-vacuum (Eppendorf, Germany) at 35$^o C$. 

\noindent{\bf 
Bulk set}
The six bulk samples of Jurkat and  U-937 cells contained 2,500 cells per sample. The cells were harvested, lysed and processed using the same procedure as for the single cells but with increased amount of trypsin and TMT labels. The samples were labeled, mixed and run as a 6-plex TMT set.

\noindent{\bf 
Mass spectrometry analysis}
Each TMT labeled set of samples was submitted for single LC-MS/MS experiment that was performed on a LTQ Orbitrap Elite (Thermo-Fisher) equipped with Waters (Milford, MA) NanoAcquity HPLC pump. Peptides were first trapped and washed onto a 5cm x 150$\mu m$ inner diameter microcapillary trapping column packed with C18 Reprosil resin (5 $\mu m$, 10 nm, Dr. Maisch GmbH, Germany). The peptides were separated on analytical column 20cm x 75 $\mu m$ of C18 TPP beads (1.8  $\mu m$, 20 nm, Waters, Milford, MA) that was heated to $60 ^oC$. Separation was achieved through applying an active gradient from $7 - 27$ \% ACN in 0.1 \% formic acid over 170 min at 200 nl/min. The active gradient was followed by a 10 min $27 - 97$ \% ACN wash step.  Electrospray ionization was enabled through applying a voltage of 1.8 kV using a home-made electrode junction at the end of the microcapillary column and sprayed from fused silica pico-tips ($20\: \mu m$ ID, 15 $\mu m$ tip end New Objective, MA). The LTQ Orbitrap Elite was operated in data-dependent mode for the mass spectrometry methods. The mass spectrometry survey scan (MS1) was performed in the Orbitrap in the range of 395 - 1,800 m/z at a resolution of $6 \times 10^4$, followed by the selection of up to twenty most intense ions (TOP20) for HCD-MS2 fragmentation in the Orbitrap using the following parameters: precursor isolation width window of 1 or 2 Th, AGC setting of 100,000, a maximum ion accumulation time of 150ms or 250ms, and $6 \times 10^4$ resolving power. Singly-charged and  and 4+ charge ion species were excluded from HCD fragmentation. Normalized collision energy was set to 37 V and an activation time of 1 ms. 
Ions in a 7.5 ppm m/z window around ions selected for MS2 were excluded from further selection for fragmentation for 20 s.

\noindent{\bf 
Analysis of raw MS data}
Raw data were searched by MaxQuant\citep{cox2008maxquant, cox2011andromeda} 1.5.7.0 against a protein sequence database including all entries from a SwissProt database and known contaminants such as human keratins and common lab contaminants. The SwissProt databases were the human SwissProt database for the U-937 and the Jurkat cells and the mouse SwissProt database for the differentiating ES cells. MaxQuant searches were performed using the standard work flow \citep{tyanova2016maxquant}. We specified  trypsin specificity and allowed for up to two missed cleavages for peptides having from 5 to 26 amino acids. Methionine oxidation (+15.99492 Da) was set as a variable modification. All peptide-spectrum-matches (PSMs) and peptides found by MaxQuant were exported in the msms.txt and the evidence.txt files. 

In addition to a standard search with the full SwissProt databases, we also searched the MS data with custom sequence databases since such searches have advantages when the sequences can be better tailored to the peptides analyzed by MS \citep{woo2013proteogenomic, nesvizhskii2014proteogenomics}. In the case of \name, we can remove sequences for lowly abundant proteins since their peptides are very unlikely to be sent for MS2. Indeed, searches with the full databases did not identify peptides from the least abundant proteins \fg{prot}a. Excluding such proteins from the search can narrow down the search space and increase the statistical power for identifying the correct peptide-spectrum-matches \citep{woo2013proteogenomic, nesvizhskii2014proteogenomics}.  To take advantage of this approach, we searched the MS data with custom databases comprised from all proteins for which MaxQuant had identified at least one peptide across many single-cell and small-bulk sets in searches against the full SwissProt databases. Theses reduced fasta databases contained 5,267 proteins for mouse and 4,961 proteins for human. Searches with them slightly increased the number of identified peptides from \name sets but such customized databases are not essential for \name.

\noindent 
The shotgun approach results in identifying different peptides in different \name sets at different levels of confidence. Because of the lower protein levels in \name sets compared to bulk sets, fewer fragment ions are detected in the MS2 spectra and thus peptide identification is more challenging than with bulk datasets. As a result, the 1\% FDR threshold that is optimal for bulk MS data may not be optimal for \name datasets.  To determine the FDR threshold that is optimal for single cell data, we plotted number of identified peptides at all levels of posterior error probability (PEP), \figs{4}. This analysis suggests that a slight increase in the arbitrary FDR threshold or $1 \%$ results in significant increase in the peptides that can be usefully analyzed across single cells while still keeping false positives low. Thus,  peptides from \name sets were filtered to 3 \% FDR computed as the mean of the posterior error probability (PEP) of all peptides below the PEP cutoff threshold \citep{Franks2016PTR}. All razor peptides were used for quantifying the proteins to which they were assigned by MaxQuant. The average number of identified peptides per TMT set  is a shown for a few \name sets in \figs{4} as a technical benchmark but it has much less practical significance than the number of  proteins that are quantified across enough single cells to be useful for analysis. This number of genes quantified across multiple sets is the standard measure for single cell RNA sequencing methods \citep{klein2015droplet}, and we have adopted it for \name as the more meaningful measure of the proteins whose levels can be analyzed across multiple single cells.

\noindent{\bf 
Data analysis }
We estimated relative peptide/protein levels from the TMT reporter ions (RI), and protein abundances from the precursor areas distributed according to the RI levels. While such estimates are well validated with bulk samples, extremely low input amounts pose unique challenges that may result in artifacts, e.g., RI intensities may reflect only background noise or the isotopic impurities of TMT tags may cross contaminate TMT channels. We evaluated the degree of background noise and found it significantly below the signal coming from the labeled peptides; see \figs{1}.  
To compensate for different amounts of total protein per channel or other channel-specific variability, the median RI intensities in each channel was set to one by diving all RI intensities by their median. In the FCAS experiment, the normalization for mCherry was performed using iRFP as a control, analogous to loading controls in western blots. After this column normalization, the vector of RI intensities for each peptide was divided by its mean or median, to remove the large differences in abundances between different peptides. The relative level of each quantified razor protein was estimated as the median of the relative levels of its peptides. All analysis relied on relative levels, i.e., the level of protein in a cell relative to its mean or median level across all cells in which the protein is quantified. Missing peptide and protein levels were imputed using the k-nearest neighbors algorithm, with k being set to one and the similarity measure for distance being the cosine of the angle between the proteome vectors.

\bibliographystyle{/Users/nslavov/GoogleDrv/B/texmf/bst/plos2015}
\bibliography{SCoPE,ribo,people/my}  


\end{spacing}

\newpage
\newcommand{\suppFig}{Supplementary } 

\newcommand{\fpa}[2]{ \begin{overpic}[width = .3\textwidth]{#1} \put(-5,70){\large \bf \hvfont  #2}\end{overpic} }
\newcommand{\inh}[2]{ \begin{overpic}[width = .28\textwidth]{#1} \put(-5,88){\large \bf \hvfont  #2}\end{overpic} }

\newcommand{\ind}[2]{ \begin{overpic}[width = .44\textwidth]{#1} \put(-3,85){\large \bf \hvfont #2}\end{overpic} }
\noindent{\Huge \suppFig Figures} \vspace{12mm} 

\begin{figure}[h!]
	 \inh{TMT_channel_noise_45C_Dist.pdf}{a}
       \hspace{0.02\textwidth}
   	 \inh{TMT_channel_noise_6plex_2.pdf}{b} 
   	   \hspace{0.02\textwidth}
   	 \inh{TMT_channel_noise_RIs_2.pdf}{c}  
\end{figure}
\noindent{\bf \suppFig Figure 1 $|$} \noise 
\newpage

\begin{figure}[h!]
   \begin{overpic}
   		[width = .4\textwidth]{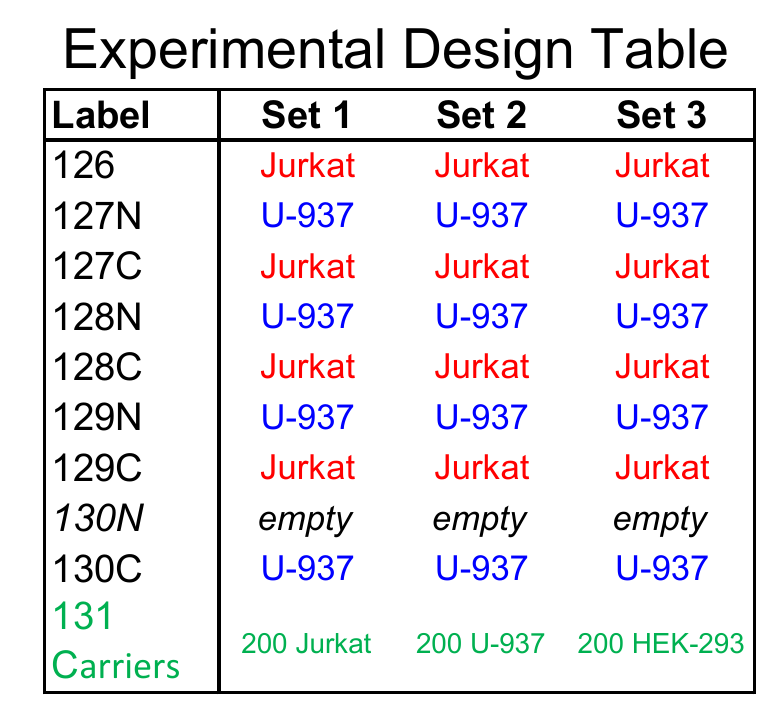} 
   		\put(-3,  92 ){\large \bf \hvfont  a}
   	\end{overpic}
   	\hspace{0.06\textwidth}
	\ind{PCA_U_J_69_Diff_Carrier.pdf}{b}\\[1em]
	\ind{Carrier_Corrs.pdf}{c}
\end{figure}
\noindent{\bf \suppFig Figure 2 $|$} \carrier_Effect
\newpage

\newcommand{\inbb}[2]{ \begin{overpic}[width = .98\textwidth]{#1} \put(0,85){\large \bf \hvfont #2}\end{overpic} }
\begin{figure}[h!]
   	 \ind{SCoPE-MS_vs_bulk_Area.pdf}{a} 
   	 \hspace{0.06\textwidth}
   	 \ind{SCoPE-MS_vs_bulk.pdf}{b} \\ [2em] 
   	 \ind{Across_37A_37A_Corrs.pdf}{c}
   	 \hspace{0.06\textwidth}
   	 \ind{38D_Reliability_vs_percentile.pdf}{d}
\end{figure}
\begin{spacing}{0.9}
\noindent{\bf \suppFig Figure 3 $|$} {\small \JUratios }
\end{spacing}
\newpage

\newcommand{\inbd}[2]{ \begin{overpic}[width = .88\textwidth]{#1} \put(0,95){\large \bf \hvfont #2}\end{overpic} }
\begin{figure}[h!]
   	 \inbd{Sorted_PEPs.pdf}{}
\end{figure}
\noindent{\bf \suppFig Figure 4 $|$} \Sorted_PEPs
\newpage

\section*{}
\subsection*{Supplementary note 1 $|$ Carrier channel influence on single cell quantification}\label{note1}
\noindent In shotgun proteomics, peptide ions sent for MS/MS are chosen based on their abundance in the MS1 survey scan, see \supp. Thus, only peptides with significant abundance in the carrier channel are likely to be sent for MS2 and quantified in the single cells. Therefore, the composition of the carrier channel can affect the sets of peptides quantified across the single cells, i.e., \name samples analyzed by a shotgun method will preferentially provide  relative quantification for proteins that are abundant in the carrier cells. However, the relative quantification of a peptide in the single cells, i.e., its RI intensities in the single cell channels, is not affected by its abundance in the carrier cells; its abundance in the carrier cells is reflected only in the 131 RI intensity. Thus changing the carrier cells in \figs{2} resulted in a slight change of the probability of quantifying some proteins, depending on the carrier cells, but hundreds of abundant proteins, such as ribosomal proteins, were quantified across all cells and carrier channels. Since, most proteins have comparable (within an order of magnitude) abundances across cell and tissue types \citep{wilhelm14,Franks2016PTR}, almost any cell type can provide useful material for the carrier channel.       
 
\vspace{2cm} 

\subsection*{Supplementary note 2 $|$ Relative quantification across \name sets}\label{note2}
\noindent \name allows quantifying only 8 cells per set (\fg{concept}), but combining multiple sets can quantify the proteomes of hundreds and thousands of cells. We were able to successfully combine relative protein levels across \name sets in two different ways: (i) When the carrier material used across sets is the same (\fg{concept}b), we used the carrier channel as a reference as established with bulk TMT samples \citep{Slavov_exp}. (ii) When the carrier material differed across carrier channels (\figs{2}), we excluded the carrier channel from the analysis and normalized the relative levels of each peptide to a mean 1 across the 8 single cells in each set, 4 Jurkat and 4 U-937 cells. Approach (ii) worked well in this case because the single cell composition cross the different \name sets was balanced. Combining \name sets based on a reference channel that is kept the same across all sets is a more versatile strategy that generalizes to any experimental design and single cell distribution across sets.

\end{document}